# Machine learning disentangles bias causes of shortwave cloud radiative effect in a climate model

Hongtao Yang[a], Guoxing Chen[a,b,*], Wei-Chyung Wang[c], Qing Bao[d], and Jiandong Li[d]

[a]Department of Atmospheric and Oceanic Sciences, Institute of Atmospheric Sciences, and CMA-FDU Joint Laboratory of Marine Meteorology, Fudan University, Shanghai, China

[b]Shanghai Frontier Science Center of Atmosphere-Ocean Interaction, Fudan University, Shanghai, China

[c]Atmospheric Sciences Research Center, University at Albany, State University of New York, Albany, NY, USA

[d]State Key Laboratory of Numerical Modeling for Atmospheric Sciences and Geophysical Fluid Dynamics, Institute of Atmospheric Physics, Chinese Academy of Sciences, Beijing, China

Corresponding author: Guoxing Chen (chenguoxing@fudan.edu.cn)

## Abstract

Large bias exists in shortwave cloud radiative effect (SWCRE) of general circulation models (GCMs), attributed mainly to the combined effect of cloud fraction and water contents, whose representations in models remain challenging. Here we show an effective machine-learning approach to dissect the individual bias of relevant cloud parameters determining SWCRE. A surrogate model for calculating SWCRE was developed based on random forest using observations and FGOALS-f3-L simulation data of cloud fraction (CFR), cloud-solar concurrence ratio (CSC), cloud liquid and ice water paths (LWP and IWP), TOA upward clear-sky solar flux (SUC), and solar zenith angle. The model, which achieves high determination coefficient > 0.96 in the validation phase, was then used to quantify SWCRE bias associated with these parameters following the partial radiation perturbation method. The global-mean SWCRE bias (in W m$^{-2}$) is contributed by CFR (+5.11), LWP (-6.58), IWP (-1.67), and CSC (+4.38), while SUC plays a minor role; the large CSC contribution highlights the importance of cloud diurnal variation. Regionally, the relative importance varies according to climate regimes. In Tropics, overestimated LWP and IWP exist over lands, while oceans exhibit underestimated CFR and CSC. In contrast, the extratropical lands and oceans have, respectively, too-small CSC and the 'too few, too bright' low-level clouds. We thus suggest that machine learning, in addition for developing GCM physical parameterizations, can also be utilized for diagnosing and understanding complex cloud-climate interactions.

**Introduction**

    Clouds play an important role in the earth climate system (e.g., refs. 1). They can reflect solar shortwave radiation back into space while trapping the longwave radiation emitted by the surface and the atmosphere. These cloud radiative effects (CREs) significantly alter the surface radiation budget and balance, affecting both the weather migration (e.g., refs. 2–3) and the climate change (e.g., refs. 4–6). However, the accurate simulation of clouds and CREs in climate models remains challenging (7) due to inadequate understanding of certain physical processes (e.g., sub-grid physics, ice nuclei; 8) and over-simplified physical parameterizations (e.g., refs. 9), casting clouds as the critical source of uncertainties in studies on climate change and climate modeling (10).
    Current climate models have marked biases in both shortwave and longwave cloud radiative effects (SWCRE and LWCRE), wherein the SWCRE bias dominates (11–13). Specifically, the models tend to simulate too weak SWCRE over the southeast Pacific (14), the Southern Ocean (15), and East Asia (16–18), allowing too much solar radiation reaching the surface. As these regions are featured with remarkable different climate regimes, the bias causes are also region dependent. For example, biases over the southeast Pacific could be partially attributed to the model deficiency in representing effects of anthropogenic aerosols on the stratocumulus microphysical properties (e.g., refs. 19–20); biases over the Southern Ocean are believed to be connected with incorrect water phase partition of mixed-phase clouds (i.e., too much ice and too little supercooled water; 21–22); and biases over East Asia are closely associated with biases in cloud diurnal cycle (i.e., too few daytime clouds and too many nighttime clouds; ref. 23) and the too-bright surface albedo over the Tibet Plateau (17). Overall, model biases in cloud micro- and macro-physical properties, and even certain non-cloud parameters (e.g., the surface albedo) may contribute to the SWCRE bias.
    While many studies have tried to isolate the causes of the SWCRE bias over various climate regimes (as discussed above), it has been difficult to quantitatively disentangle the contributions of diverse factors. One important reason is that the intricate framework of climate models, replete with too-many elements and feedback processes (some of which could be artificial rather than physical-based), renders the models akin to 'black boxes' and difficult to interpret the results clearly for most users (24). One the other hand, most models are computationally expensive, hindering the users from conducting enough sensitivity simulations for fully understanding the model responses to perturbations in model parameters. To overcome these problems, previous studies usually followed two methods: the radiative kernel method (e.g., refs. 25–27) and the partial radiation perturbation (PRP) method (e.g., refs. 25, 28–31). While both methods are effective, they are still computationally expensive to some degree.
    In this study, we propose a machine-learning based method to quantitatively disentangle bias causes for the SWCRE bias in the FGOALS-f3-L model, a general circulation model developed by the Institute of Atmospheric Physics, Chinese Academy of Sciences (32–36). The method is to first build a surrogate model using the random

forest (RF; 37–38) to emulate the radiation calculation, and then calculate the contribution of biases in each SWCRE-related cloud property following the PRP approach using the surrogate model.

**Results**

**Spatial distributions of model biases**

Figure 1 presents the spatial distributions of the observed SWCRE, CFR, CSC, LWP, IWP, and SUC. SWCRE is generally negative globally, having large values over the Intertropical Convergence Zone (ITCZ), East Asia, the storm-track regions, the eastern subtropical oceans, and the Southern Ocean (Fig. 1A). This pattern is very similar to that of CFR (Fig. 1C), suggesting the dominant role of CFR. CSC exhibits a marked land-ocean contrast, with large values over land and small values over the oceans (Fig. 1E).

LWP shows large values over East Asia, the storm-track regions, the eastern subtropical oceans, and the Southern Ocean (Fig. 1G), while IWP has large values over the ITCZ and the storm-track regions (Fig. 1I). The different spatial distributions of LWP and IWP and their correspondences with the CFR distribution indicate that the cloud type varies remarkably across different climate regimes.

SUC also presents a land-ocean contrast, showing large values over land and small values over the oceans (Fig. 1K). This is because the land regions usually have larger surface albedo (e.g., the deserts and the Tibet Plateau) and larger aerosol loading (e.g., East Asia) than the oceanic regions.

Also shown in Fig. 1 are the model biases in SWCRE, CFR, CSC, LWP, IWP, and SUC (model minus observation). The pattern of SWCRE bias is very similar to the CMIP5/6 multi-model ensemble mean (MME) results given in previous studies (11–12, 23), showing significant underestimation over East Asia, the eastern subtropic oceans, and the Southern Ocean, and overestimation over Africa, Australia, South and North America (Fig. 1B). The pattern of CFR bias is very similar to that of SWCRE, but with the opposite phases, i.e., regions with overestimated (underestimated) CFR tend to have too strong (weak) SWCRE (Fig. 1D vs. Fig. 1B). This is consistent with dominant role of CFR in determining the SWCRE. Nevertheless, some exceptions are noticed. For example, the tropical Africa shows underestimated CFR but overestimated SWCRE. CSC is underestimated over most regions, notably the land area. This bias pattern is also similar to the MME results shown in G. Chen, et al. (23).

The biases of LWP and IWP presents similar patterns, showing overestimations over most regions (Fig. 1H vs. Fig. 1J). It is noticed that the largest overestimations are located over the ITCZ and the storm-track regions, whereas that the model tends to simulate too small CFR and too weak SWCRE over these regions. This indicates that the contributions of LWP/IWP biases and CFR bias to the SWCRE bias could be compensative to each other, causing difficulties for the model diagnosis and possible model improvement.

The model overestimates SUC over most areas, indicating the model simulates larger planetary albedo. The model bias over the oceans could be due to that the model simulated too-high relative humidity (39) while the bias over regions around the Tibet Plateau should be attributed to the too-bright surface albedo (17 and 40).

**Individual contributions to the SWCRE bias**

First of all, it is noted that the RF regression of SWCRE is remarkably successful. It reaches very large coefficients of determination over both the observation (0.96) and simulation data (0.98). Meanwhile, the total SWCRE bias calculated with the PRP method (i.e., the sum of $B_{CFR}$, $B_{CSC}$, $B_{LWP}$, $B_{IWP}$, and $B_{SUC}$) is also very close to the actual SWCRE bias. Shown in Fig. S2, the PRP-estimated SWCRE (Fig. S1A) bias has a spatial correlation coefficient of 0.96 with the actual SWCRE bias (Fig. 1B). The residual is small throughout the domain (Fig. S1B), with the domain mean of -0.40 W m$^{-2}$. These indicate that the RF surrogate model and the PRP method are trustworthy for the bias decomposition.

Figure 2 presents the spatial distributions of individual contributions of the model bias in CFR, CSC, LWP, IWP, and SUC to the SWCRE bias. Four features are worthy of highlighting. First, the estimated bias contributions have the expected signs (i.e., consistent with the qualitative analysis; for example, a larger CFR/LWP should cause a stronger SWCRE, and vice versa) over most regions (highlighted with dots in the figure) for all five variables. This further confirms that the RF surrogate model and the PRP-based decomposition are reasonable in the physics.

Second, the cloud diurnal variation, which has been ignored in most model diagnosis studies, contributes greatly to the SWCRE bias. Its domain-mean contribution is 4.38 W m$^{-2}$, close to the that of the daily-mean cloud fraction (5.11 W m$^{-2}$). However, the spatial patterns of $B_{CFR}$ and $B_{CSC}$ differ clearly. The former is positive over the oceans and the tropical land but negative over the extratropical land, while the latter is positive over most regions, consistent with the simulated too-small CSC as shown in Fig. 1F. This means that, if excluding CSC from the bias diagnosis, CFR may not be able to fully absorb the bias contribution by CSC and cause incorrect bias decomposition (this will be discussed later), misleading the possible model improvement.

Third, the bias contributions of LWP and IWP are compensating those by CFR and CSC. Although the overestimation in LWP is smaller than that in IWP, the LWP bias causes larger SWCRE bias, with the domain mean of -6.58 vs. -1.67 W m$^{-2}$ by the IWP bias. This is because the solar reflectance of liquid water is much larger than that of ice water.

Fourth, the bias contribution of SUC is generally small and does not have expected sign over vast areas. Its domain mean is -0.35 W m$^{-2}$, closer to the domain-mean residual. Its values are only marked over the Tibet Plateau, where the regional mean is about 5 W m$^{-2}$.

Figure 3 presents the zonal-mean characteristics of the decomposed SWCRE bias. Over land, $B_{LWP}$ dominates the overestimated SWCRE, especially for the tropics; $B_{CSC}$ is smaller than $B_{CFR}$ at most latitudes; and the sum of $B_{CFR}$ and $B_{CSC}$ can be mostly canceled by $B_{IWP}$. Over the oceans, $B_{CFR}$ is larger than $B_{CSC}$ and dominates the underestimated SWCRE at most latitudes; $B_{LWP}$ is still the largest over the tropics, but much smaller than its counterpart over land.

To get more details, the bias decomposition was further analyzed using the KMeans clustering method (see description in Fig. S2 in the supplementary materials). Shown in Fig. 4, this approach identified four distinct climate regimes: (I) the tropical land areas; (II) the extratropical land areas excluding East Asia; (III) deep-convection areas consisting of the ITCZ, the storm-track regions, and East Asia; and (IV) shallow-convection ocean aeras.

Therein, regime I is dominated by the overestimated LWP and IWP, leading to too strong SWCRE that cannot be offset by the underestimated CFR and CSC. Regime II is dominated by the underestimated CSC, whose effect can be largely balanced by those from the overestimated CFR, LWP, and IWP. Regime III mainly suffer from the too small CFR and too small CSC. Regime IV, where ice clouds seldom occur in both the observation and simulations, is clearly dominated by the 'too few, too bright' issue of low-level clouds, similar to many other GCMs (41).

These findings offer valuable insights for model improvement. For example, the overestimation of cloud water paths is linked with the Resolving Convective Precipitation scheme employed in the FGOLA-f3-L model, which explicitly calculates convective precipitation using the cloud microphysical scheme (35). Implementing a regime-dependent adjustment or tuning algorithm should help mitigate the associated bias. Meanwhile, considering the extensive coverage and low albedo of the underlying ocean surface in regime IV, it is concluded that improving the representation of low-level clouds in the model could significantly reduce the modeled SWCRE bias.

**Discussion**

This study for the first time quantitatively estimates individual contributions of the model biases in cloud (CFR, CSC, LWP, and IWP) and non-cloud (SUC) properties to the SWCRE bias, a longstanding issue that has been annoying the community in studies on climate modeling and climate change. Three points are highlighted.

First, it is evident that the compensation among individual biases is common in all climate regime, as further illustrated in Fig. S3 in the supplementary materials. These compensations arise from both physical (e.g., shortcomings of physical parameterizations; 42–43) and non-physical (e.g., model tuning strategies; 44) factors. The insights gained from this study are instrumental for understanding the interplay of factors related to the SWCRE, and more importantly, provide a framework for improving/fine-tuning the climate model, taking into consideration both the overall energy balance and the interactions between various cloud variables.

Second, the modeling of cloud diurnal variation calls for more attention in the model diagnosis and development. Its contribution to the SWCRE bias is comparable to that of the daily-mean cloud fraction and cannot be absorbed by the latter in most cases. Excluding CSC from the surrogate model not only reduces the model physical rationality but also distorts the bias decomposition because of the compensation effect (shown in Fig. S4 in the supplementary materials), which would mislead efforts of model improvement and tuning.

Third, the RF-based surrogate model demonstrates great effectiveness and efficiency. It not only saves computational resources but also reduces the difficulties of involving new parameters in building a model (e.g., it is difficult to account for cloud diurnal variation in the conventional radiative-kernel and PRP models). Thus, the approach is inherently suitable for understanding processes that involves complex interactions (e.g., the surface-air interactions; 45) in both the observation and modeling realms. This facilitates better diagnosing GCMs for model improvement besides developing physical parameterizations.

**Materials and Methods**

**Surrogate radiation model based on random forest**

We construct an RF-based radiation surrogate model to emulate the relationship between SWCRE at the top of the atmosphere (TOA) and the associated cloud and non-cloud atmospheric variables. RF is a supervised machine-learning algorithm that fits multiple decision trees and combines their outputs to reach a single result. It is handy in building data-driven prediction models and has demonstrated great performances in many atmospheric studies, such as estimating aerosol optical depth (46–47), parameterizing the sub-grid model physics in a climate model (48), and diagnosing aerosol-cloud interactions (49–50).

SWCRE is defined as the disparity between all-sky and clear-sky shortwave radiative fluxes (all-sky minus clear-sky; ref. 51). Therefore, six SWCRE-related variables were taken as predictors in building the surrogate model: the daily-mean cloud fraction (CFR), the cloud-solar concurrence (CSC) ratio, the cloud liquid and ice water paths (LWP and IWP), the upward shortwave flux at the TOA over the clear sky (SUC), and the cosine of solar zenith angle (CSZ) at the given grid.

Therein, CSC is a metric proposed by Chen and Wang (52) and G. Chen, et al. (23) to quantitatively measure cloud diurnal variation. It is defined as the ratio of the effective-day cloud fraction, which is the solar-weighted daily-mean cloud fraction (19), to the conventional daily-mean cloud fraction (i.e., CFR). Thus, it considers the concurrence probability of clouds and solar radiation. A larger (smaller) CSC indicates that the clouds tend to occur near (away from) the local noontime or during the daytime (nighttime), and should yield larger (smaller) SWCRE given the same CFR.

LWP and IWP are the proxies for cloud optical depth.

SUC is to serve as a proxy for the clear-sky planetary albedo, including effects of both the surface albedo and the planetary albedo changes caused by aerosols aloft in the atmosphere.

CSZ is to account for the seasonal and latitudinal variations of the incoming solar radiation at the TOA. It is calculated as

$$CSZ = \max(\cos(\phi - \phi_s), 0) \qquad (1),$$

where $\varphi$ and $\varphi_s$ indicate the latitudes of the given grid and the subsolar point, respectively.

Aerosol-cloud interactions were not considered in building the surrogate model. There are mainly two reasons. First, there is not a reliable observational dataset that provides global information of aerosol compositions and concentrations; and second, the treatment of aerosol-cloud interactions was not very realistic in the FGOALS-f3-L model, where aerosols were prescribed rather than fully simulated. Nevertheless, as aerosol-cloud interactions may cause changes in cloud cover and cloud thickness, it is believed that CFR, LWP, and IWP should carry some information of aerosol-cloud interactions.

Both observational and simulation data were used to build the surrogate model. For the observation, we obtained SWCRE and CFR from the Clouds and the Earth's Radiant Energy System (CERES) project — CERES_SYN1deg_Ed4A (2005–2015) dataset (53), LWP and IWP from the Moderate-resolution Imaging Spectroradiometer (MODIS) product — MCD06COSP_M3_MODIS (2005–2015) dataset (54), and calculated CSC using 3-hourly cloud fraction data from the International Satellite Cloud Climatology Project (ISCCP)-H (1984–2014) dataset (55). For the simulation, all the above variables were obtained from the historical simulation (1980–2014) conducted by the FGOALS-f3-L model, included in the sixth Coupled Model Intercomparison Project (CMIP6).

The multi-year averaged monthly means were calculated for all variables, and both the observation and simulation data were regridded to the resolution of 1° latitude × 1° longitude to facilitate observation-model intercomparisons. The analysis was limited to regions between 60°S–60°N which have solar insolation throughout the year and relatively-more reliable satellite observations. The data at each grid of a month are treated as a sample, and thus the observation and simulation data both have 360 × 120 × 12 (~ 0.5 million) samples.

We randomly drew 70% samples from both the observation and simulation data and mixed them to make the training dataset, and took the rest 30% as the test dataset. Because the simulation data have different statistical characteristics than the observation data (e.g., the simulate data generally has larger LWP and IWP but smaller CFR than the observation, which is shown in the main text), the mixing use of simulation and observation data in training can increase the model generalizability. The number of trees in RF was set to 1000. Using more or fewer trees has little effect on results in the main text.

The resulting model well predicts SWCRE in both the observation and simulation. The coefficients of determination (i.e., $R^2$) over the test data are 0.96 and 0.98 for the observation and simulation, respectively. The relative importance of six predictors from high to low is CFR (0.29), SUC (0.28), LWP (0.20), IWP (0.11), CSC (0.07), and CSZ (0.06).

**Decomposition of the SWCRE bias**

The surrogate model is taken as an implicit function,

$$y = f(x_i), \ i=1, 2, 3, 4, 5, 6 \tag{2}$$

where $y$ is SWCRE while $x_i$ refers to CFR, CSC, LWP, IWP, SUC, and CSZ. Then, following the PRP method, the SWCRE changes caused by changes in a certain predictor, for example, $x_1$, can be calculated as

$$\Delta y_{x_1} = \frac{\partial f}{\partial x_1} \Delta x_1 + O(\Delta x_1)$$

$$\approx f(x_{1,s}, x_2, ..., x_6) - f(x_{1,o}, x_2, ..., x_6) \tag{3}$$

where the subscript $s$ and $o$ indicate simulation and observation, respectively. Because of the nonlinear nature of radiation transfer, the estimated $\Delta y_{x1}$ using Eq. (3) would be very sensitive to values of $x_2, \ldots, x_6$. Hence, we used the mean results of the estimates at $N+1$ points to improve the calculation accuracy,

$$\Delta y_{x_1} = \frac{1}{N+1} \sum_{j=0}^{N} \left[ f(x_{1,s}, x_2 + \frac{j}{N}\Delta x_2, ..., x_6 + \frac{j}{N}\Delta x_6) \right.$$
$$\left. - f(x_{1,o}, x_2 + \frac{j}{N}\Delta x_2, ..., x_6 + \frac{j}{N}\Delta x_6) \right] \quad (4),$$

where
$$\Delta x_i = x_{i,s} - x_{i,o} \quad (5).$$

Setting $x_1$ to one of CFR, CSC, LWP, IWP, and SUC, and $x_2$–$x_6$ to the rest 5 variables, we calculated the individual contributions of CFR, CSC, LWP, IWP, and SUC to the modeled SWCRE biases following Eq. (4). The modeled CSZ has no biases and thus no contributions to the SWCRE bias. These individual bias contributions are termed as $B_{CFR}$, $B_{CSC}$, $B_{LWP}$, $B_{IWP}$, and $B_{SUC}$ in the manuscript, respectively. $N$ was set to 4 in this study. The use of larger $N$ has been tested and does not cause much changes to the results.

## Acknowledgments

This study is supported by the National Natural Science Foundation of China (42275074).

## Data availability

The CERES data are available at https://ceres-tool.larc.nasa.gov/ord-tool/jsp/SYN1degEd41Selection.jsp; the MODIS data are available at https://ladsweb.modaps.eosdis.nasa.gov/; the ISCCP-H data are available at https://www.ncei.noaa.gov/products/international-satellite-cloud-climatology; and the FGOALS-f3-L model data are available at https://aims2.llnl.gov/search/cmip6/.

**Figures**

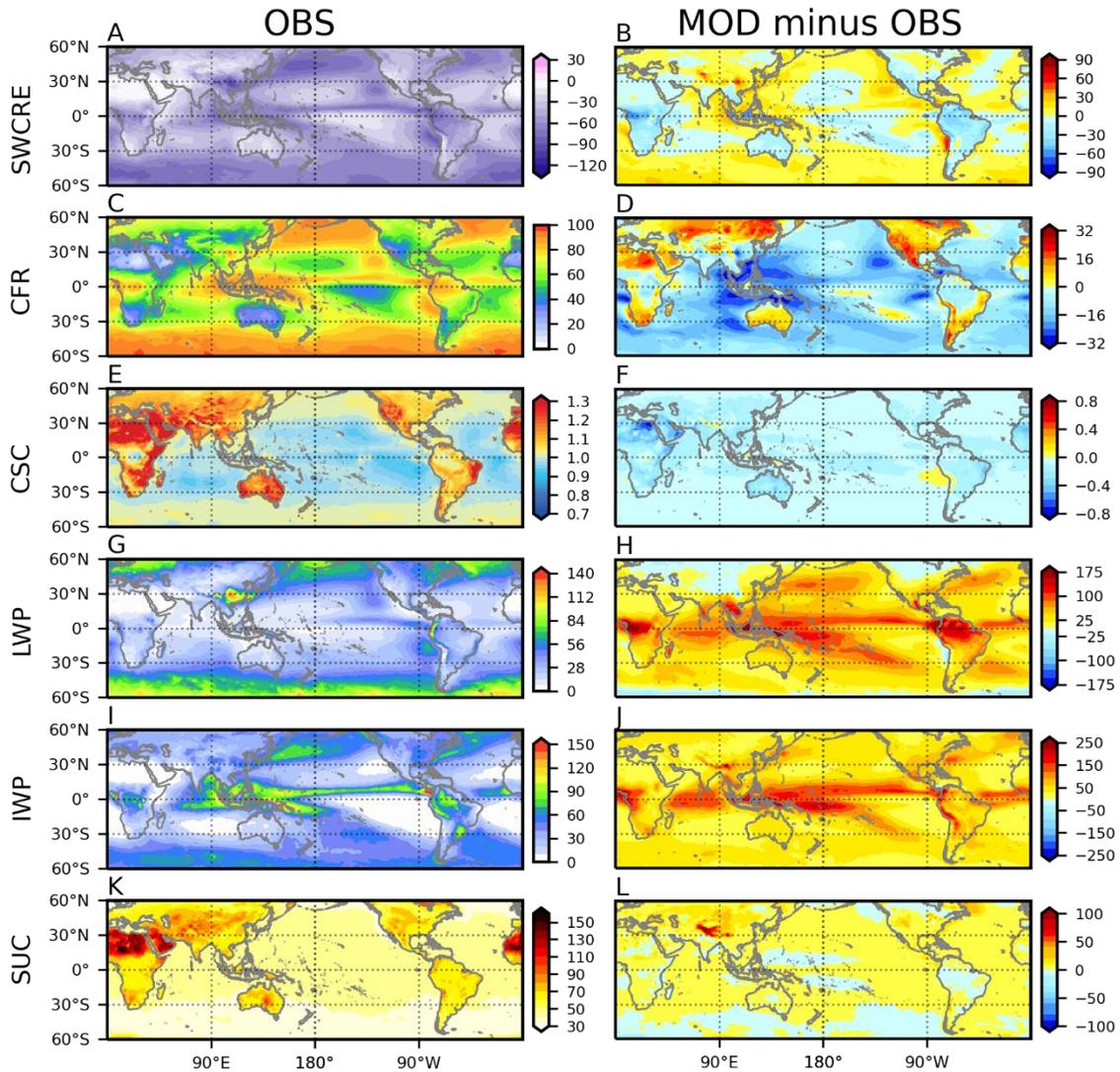

**Figure 1.** Model-observation comparisons in the annual-mean spatial distribution of shortwave cloud radiative effect (SWCRE, W m$^{-2}$; A and B), the cloud fraction (CFR, %; C and D), the cloud-solar concurrence ratio (CSC , E and F), the cloud liquid water path (LWP, g m$^{-2}$; G and H), the cloud ice water path (IWP, g m$^{-2}$; I and J), and the upward shortwave flux over the clear sky (SUC, W m$^{-2}$; K and L): (left) observations; and (right) the model biases (model minus observation). SWCRE and SUC were both calculated at the top of the atmosphere.

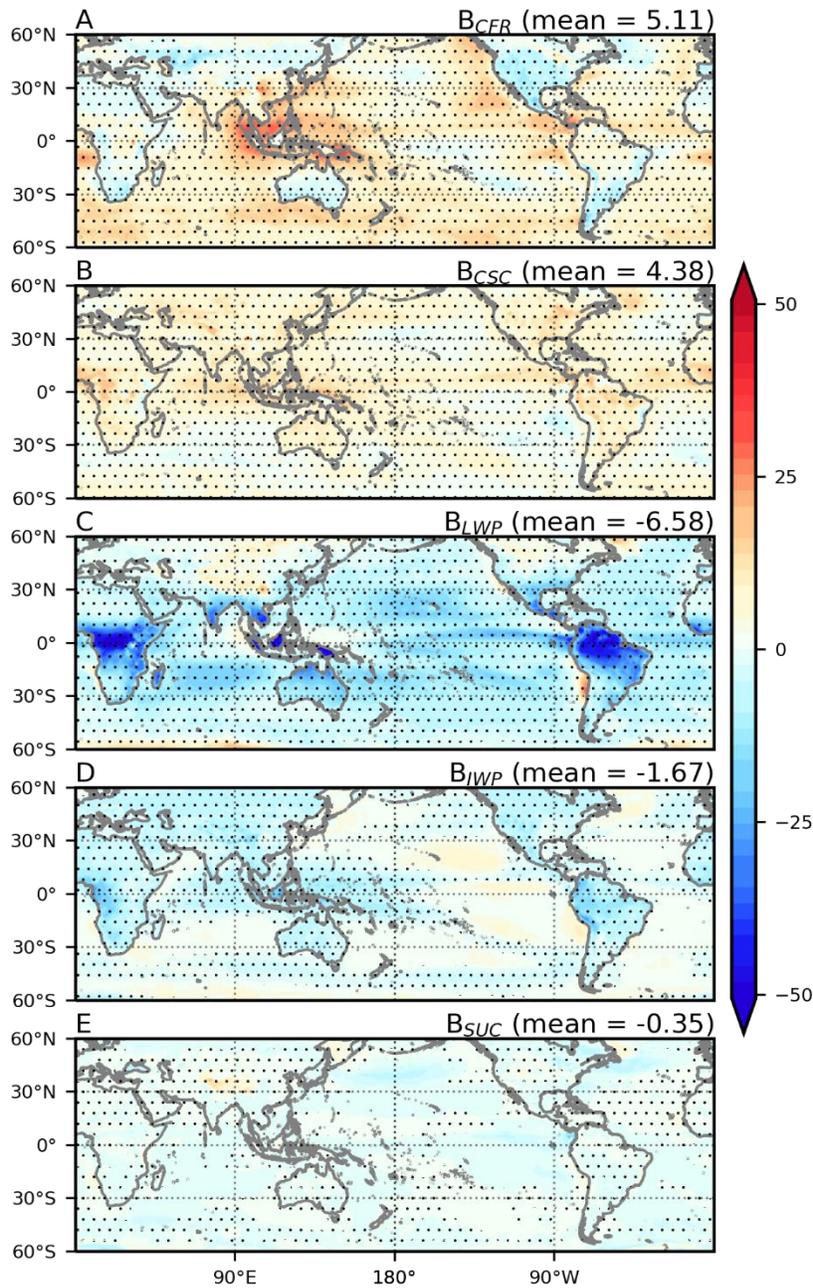

**Figure 2.** Contributions of the model biases in CFR (A), CSC (B), LWP (C), IWP (D), and SUC (E) to the SWCRE bias (W m$^{-2}$). The numbers at upper-right corners indicate the domain means. Regions where the estimated bias contributions have expected signs are highlighted with dots. The modeled solar zenith angle has no bias and thus no contribution to the SWCRE bias.

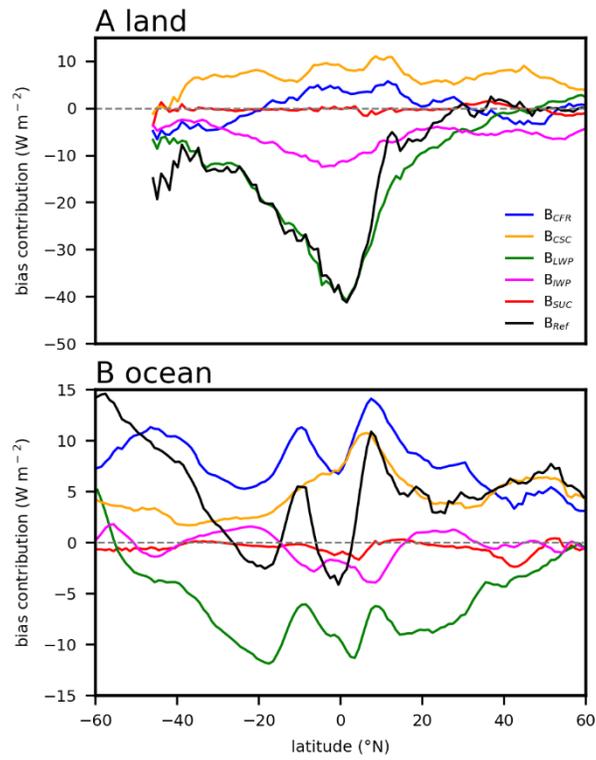

**Figure 3.** Zonal-averaged contributions of the model biases in CFR, CSC, LWP, IWP, and SUC to the model biases in SWCRE (W m$^{-2}$) over land (A) and oceans (B). The solid black lines ($B_{Ref}$) indicate the total SWCRE biases, calculated by the modeled SWCRE minus the observed one.

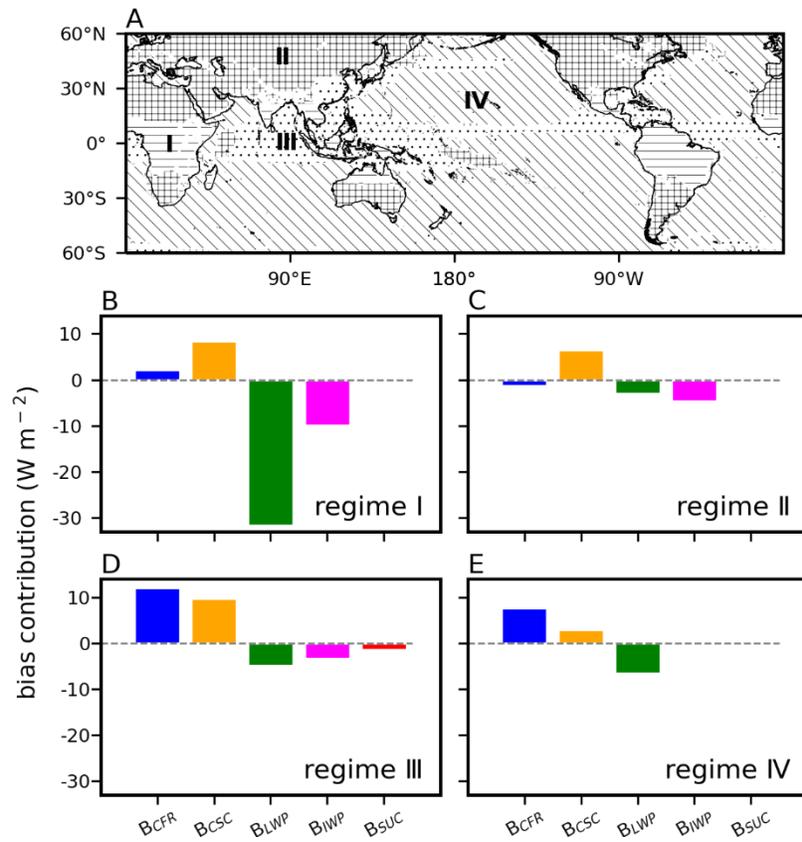

**Figure 4.** Spatial distribution of bias regimes identified through the KMeans clustering (A) and the composite characteristics of bias decompositions over individual regimes (B–E).

**Supplementary figures**

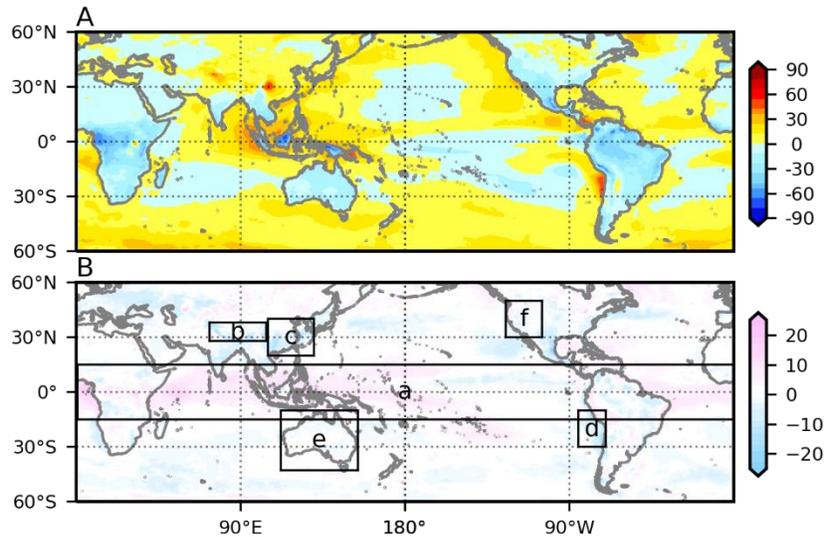

**Fig. S1.** Total SWCRE bias calculated with the PRP method (i.e., the sum of $B_{CFR}$, $B_{CSC}$, $B_{LWP}$, $B_{IWP}$, and $B_{SUC}$; W m$^{-2}$; A) and the residual (i.e., the actual SWCRE bias minus the PRP-estimated SWCRE bias; W m$^{-2}$; B). The spatial correlation coefficient between the PRP-estimated and actual SWCRE biases is 0.96, and the domain-mean residual is -0.40 W m$^{-2}$. Black boxes in B indicate selected regions where results are further examined in Fig. S3: a. 15 °S–15 °N (ocean only); b. 28 °S–38 °N, 73 °E–104 °E; c. 20 °N–40 °N, 105 °E–130 °E; d. 30 °S–10 °S and 85 °W–70 °W (ocean only); e. 43 °S–10 °S and 112 °E–154 °E (land only); and f. 30 °N–50 °N and 125 °W–105 °W (land only).

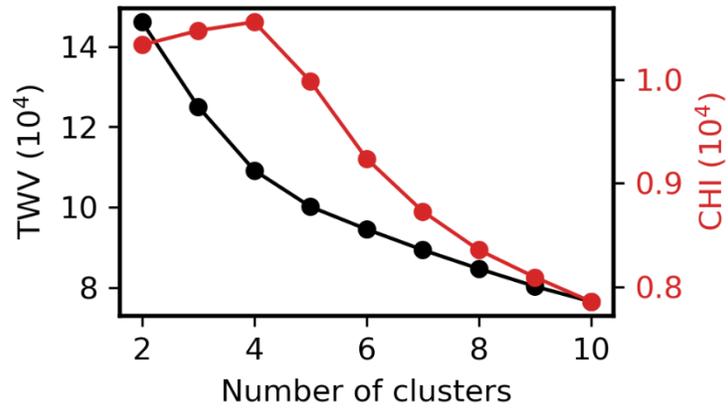

**Fig. S2.** Total within-cluster variance (TWV, with respect to the left y axis; black line) and Caliński Harabasz index (CHI, with respect to the right y axis; green line) as a function of cluster number in the KMeans clustering. The KMeans method categorizes samples into a predetermined number of clusters where each sample belongs to the cluster whose center has the shortest distance to the sample. The optimal cluster number is determined as the number when adding another cluster does not much decrease the TWV or the number where the CHI reaches the maximum. Both TWV and CHI results indicate that the optical cluster number is 4 in this study.

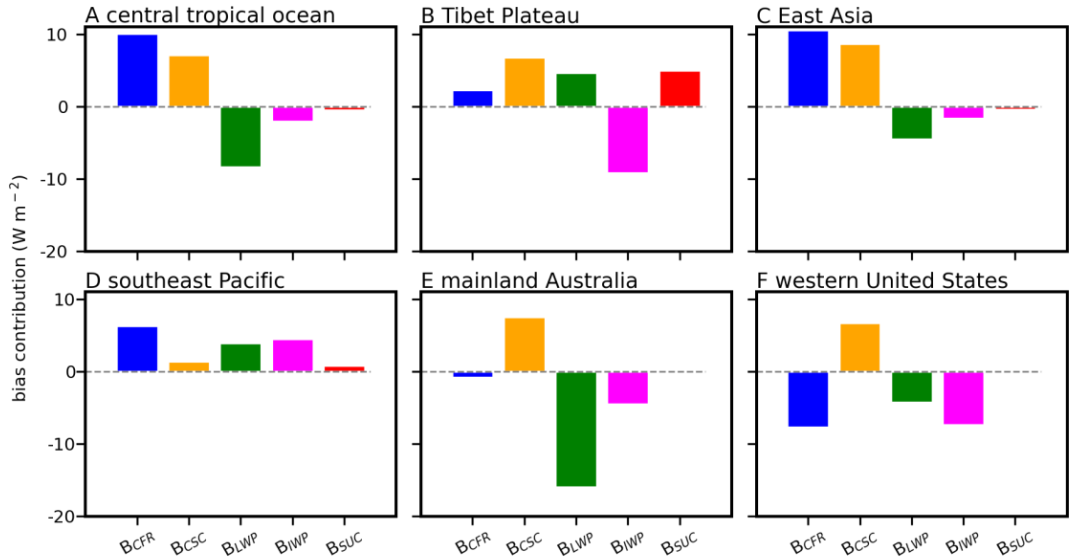

**Fig. S3.** Contributions of the model biases in CFR, SUC, LWP, IWP, and CSC to the modeled SWCRE biases over the central tropical ocean (A), the Tibet Plateau (B), East Asia (C), the southeast Pacific (D), the mainland Australia (E), and the western United States (F). The ranges of these regions are shown in Fig. S1. The compensation widely exists among different terms: the commonly-aware compensation between CFR and LWP/IWP over the central tropical ocean, the Tibet Plateau, and East Asia, the compensation between CSC and LWP/IWP over the Tibet Plateau, East Asia, mainland Australia, and western United States, and the compensation between CFR and CSC over the mainland Australia and western United States. Specifically for the western United States, our previous study (1) has shown that the regional SWCRE bias could be increased if we corrected the CSC alone. Over the southeast Pacific, all five terms appear to cause weaker SWCRE. However, it is an illusion of the annual mean. When we check the decomposition in each month, it is clear that the compensation is evident in December–February, when the simulated CFR, CSC, and LWP causes stronger SWCRE while the simulated IWP and SUC causes weaker SWCRE (figure not shown).

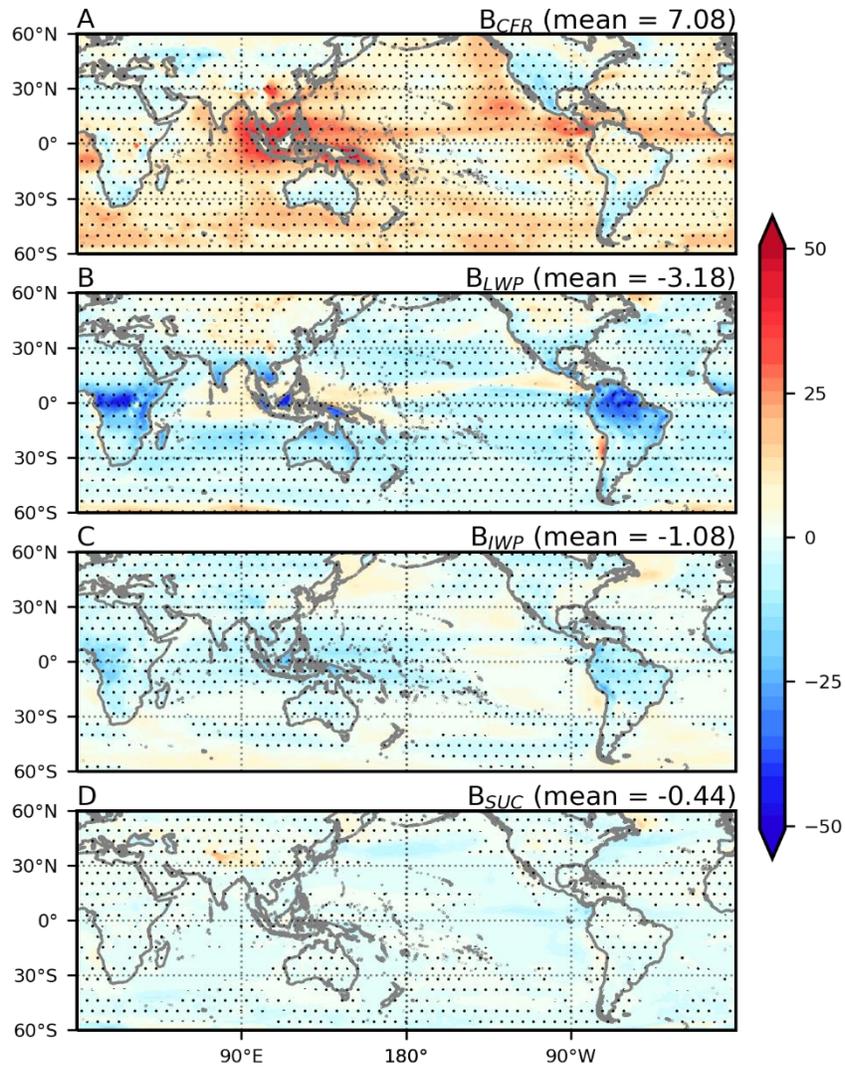

**Fig. S4.** Contributions of the model biases in CFR (A), LWP (B), IWP (C), and SUC (D) to the SWCRE bias (W m$^{-2}$) when CSC is not considered in the surrogate model. The numbers at upper-right corners indicate the domain means. Regions where the estimated bias contributions have expected signs are highlighted with dots. Excluding CSC greatly reduces the physical rationality of the surrogate model (e.g., clearly less dots in the plots) and distorts the bias decomposition because of the compensation effect.